\documentclass{article}
\usepackage{spconf,amsmath,graphicx}
\usepackage[colorlinks, linkcolor=red,anchorcolor=red, citecolor=green]{hyperref}
\usepackage{graphicx}
\usepackage{url}
\usepackage{booktabs}
\usepackage{multirow}
\usepackage{float}
\usepackage{color}
\title{Universal Learned Image Compression with Low Computational Cost}
\name{Bowen Li$^{1}$, Yao Xin$^{2}$, Youneng Bao$^{1}$, Fanyang Meng$^{2}$, Yongsheng Liang$^{1}$\sthanks{Corresponding\ author,\ email: liangys@hit.edu.cn}, Wen Tan$^{1}$}
\address{$^{1}$Harbin Institute of Technology, Shenzhen, China\ \ \ \ \ $^{2}$Peng Cheng Laboratory, Shenzhen, China}

\begin{document}
%\ninept
%
\maketitle
\begin{abstract}
Recently, learned image compression methods have developed rapidly and exhibited excellent rate-distortion performance when compared to traditional standards, such as JPEG, JPEG2000 and BPG. However, the learning-based methods suffer from high computational costs, which is not beneficial for deployment on devices with limited resources. To this end, we propose shift-addition parallel modules (SAPMs), including SAPM-E for the encoder and SAPM-D for the decoder, to largely reduce the energy consumption. To be specific, they can be taken as plug-and-play components to upgrade existing CNN-based architectures, where the shift branch is used to extract large-grained features as compared to small-grained features learned by the addition branch. Furthermore, we thoroughly analyze the probability distribution of latent representations and propose to use Laplace Mixture Likelihoods for more accurate entropy estimation. Experimental results demonstrate that the proposed methods can achieve comparable or even better performance on both PSNR and MS-SSIM metrics to that of the convolutional counterpart with an about 2\(\times\) energy reduction.
\end{abstract}
\section{Introduction}
\label{sec:Introduction}

Image compression is one of the most fundamental fields in signal processing, which aims to reach a trade-off between bitrate and distortion. With the rapid development of deep learning, learned image compression methods \cite{Balle2017,Balle2018,Minnen2018a,Cheng2020,Xie2021} have drawn much attention and exhibited their outstanding rate-distortion performance when compared to the traditional standards, such as JPEG \cite{JPEG}, JPEG2000 \cite{JPEG2000} and BPG \cite{BPG}. Although numerous progress has been made, these learning-based methods suffer from large computational complexity, which is not conducive to deployment on devices with limited resources. Therefore, it is necessary to propose a novel image compression framework with low computational costs while maintaining comparable rate-distortion performance.
\begin{figure}[ht]  
\centering  
\includegraphics[width=8.4cm]{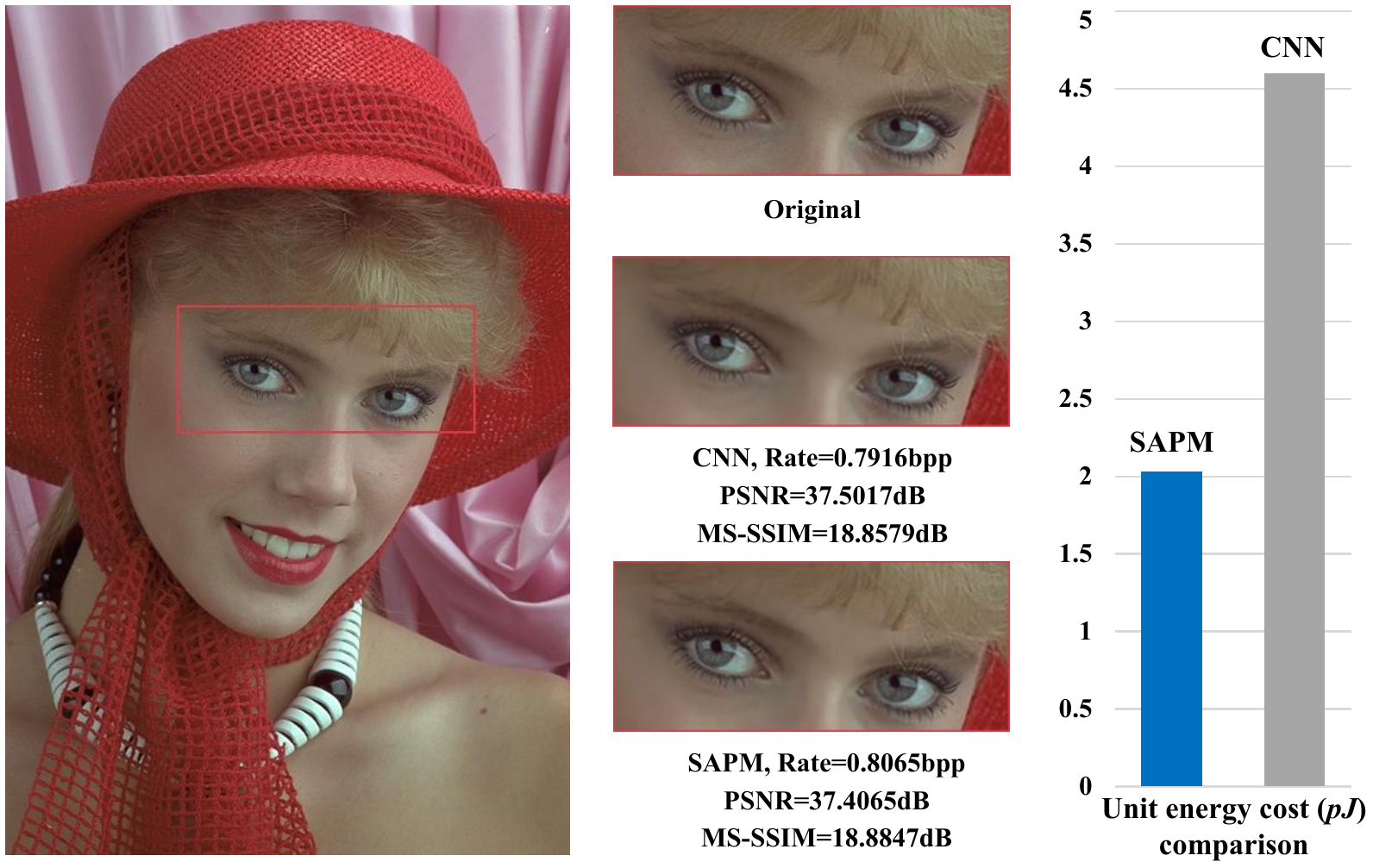}
\vspace{-0.3cm}
\caption{Visualization of reconstructed images \textit{Kodim04} from Kodak dataset and unit energy cost comparison.} 
\label{visualization}
\vspace{-0.5cm}
\end{figure}

To solve the problem, several methods have been developed. One of the most popular methods is pruning \cite{Prunning}, which aims to remove redundant or less important weights to compress and accelerate the original network. Another popular method is knowledge distillation \cite{KD}, and the performance of lightweight student networks can be improved through the knowledge conveyed by the corresponding teacher networks. However, these compressed models still contain massive multiplication operations and consume a lot of energy, and do not benefit deployment. Although there exist model quantization methods \cite{BNN} that can reduce energy consumption, they cannot reach comparable performance when compared to their baselines. Recently, Chen \textit{et al.} \cite{AdderNet} pioneered AdderNet, which utilizes the \(L1\) norm to calculate the similarities between inputs and filters. This novel operation can avoid multiplication and reduce massive energy consumption. In the same year, Elhoushi \textit{et al.} \cite{deepshift} proposed DeepShift, which quantizes each weight of convolutional networks to a power of 2. In the aspect of hardware, applying bit-shift operation on an element is equivalent to multiplying it by a power of 2, and it means that DeepShift can realize bit-shift operation through special quantization. Thus, both AdderNet and DeepShift are suitable for deployment to resource-constrained devices, such as mobile phones and embedded devices.

Recently, Li \textit{et al.} \cite{9747652} have proposed AdderIC, which utilizes AdderNet to construct an image compression framework. However, it still exists a performance gap between AdderIC and its CNN counterpart. To this end, in this paper, we try to make full use of both AdderNet and DeepShift to propose a novel image compression framework with low computational costs and comparable rate-distortion performance. Our main contributions are as follows:
	
\(\bullet\) We propose shift-addition parallel modules (SAPMs), including SPAM-E for the encoder and SAPM-D for the decoder, to largely reduce energy consumption and facilitate deployment on those computing-constrained devices. Besides, the proposed SAPMs are plug-and-play components, which means that they have strong generalization ability and can be applied in other CNN-based architectures.

\(\bullet\) We thoroughly analyze the probability distribution of latent representations and propose a Laplace Mixture Model for more accurate entropy estimation.

\(\bullet\) Experimental results show that the proposed methods can bring about 2\(\times\) energy reduction while maintaining comparable or even better rate-distortion performance on both PSNR and MS-SSIM metrics as shown in Fig. \ref{visualization}.

\section{Proposed Method}
\label{sec:Proposed Methods}
\subsection{Motivation}
From \cite{energy, dally2015high, shiftaddnet}, we can get energy consumption among different operations and formats when implemented in a 45nm CMOS technology as follow:
\vspace{-0.3cm}
\begin{table}[h]
\begin{center}
\vspace{-0.2cm}
 \caption{Energy consumption comparison}
 \label{table1}
 \begin{tabular}{llcr}
	\toprule
		Operation&Format&Energy Cost (\textit{pJ})&Improv.\\
	\midrule
		\# Mult.&FP32&3.70&-\\
        \multirow{2}{*}{\# Add.} &FP32&0.90&\textbf{4.1\(\times\)}\\
            &FIX32&0.10&\textbf{37\(\times\)}\\
        \# Shift &FIX32&0.13&\textbf{28.5\(\times\)}\\
	\bottomrule
    \vspace{-0.9cm}
\end{tabular}
 \end{center}
 \end{table}
 
From Tab. \ref{table1}, both shift and addition can bring a cheap operation, while multiplication suffers from extremely large energy consumption. Besides, through experimental observation, neither shift nor addition can achieve similar performance as multiplication-based networks (\textit{e.g.}, CNN). To this end, we try to combine these two weak players and develop low-power shift-addition parallel modules (SAPMs) for image compression, where the shift branch can be used to extract large-grained features and the addition branch can learn small-grained features.

\subsection{Shift-Addition Parallel Modules}
\subsubsection{Shift Branch}
DeepShift was first proposed in \cite{deepshift}, and it adopts bitwise shift operation to construct networks. Specifically, there are two methods to train the DeepShift model, including DeepShift-Q and DeepShift-PS. DeepShift-Q quantizes the weight \(W\) in networks to \(W_{q}\) by rounding it to the nearest power of 2, while DeepShift-PS trains shift and sign parameters directly. In this paper, DeepShift-Q is adopted to train our shift branch and the forward pass as follow:
\vspace{-0.1cm}
\begin{equation}
S=\operatorname{sign}(W), P=\log _{2}(\operatorname{abs}(W)), W_{q}=S \cdot 2^{p}
\vspace{-0.1cm}
\end{equation}
where \(S\) denotes sign matrix and \(P\) represents shift matrix. Once quantized, it can realize shift operation since applying bitwise shift on an element is equivalent to multiplying it by a power of 2 in hardware. 

Although bitwise shift operation exhibits its cheap cost in terms of energy consumption, it cannot achieve the same or similar expressive capacity as its original counterpart. This is because that shift branch will ignore some important information during quantization, which means it can merely be used to extract coarse-grained features.
% \vspace{-0.8cm}
\begin{figure}[t]  
\centering  
\includegraphics[width=8.4cm]{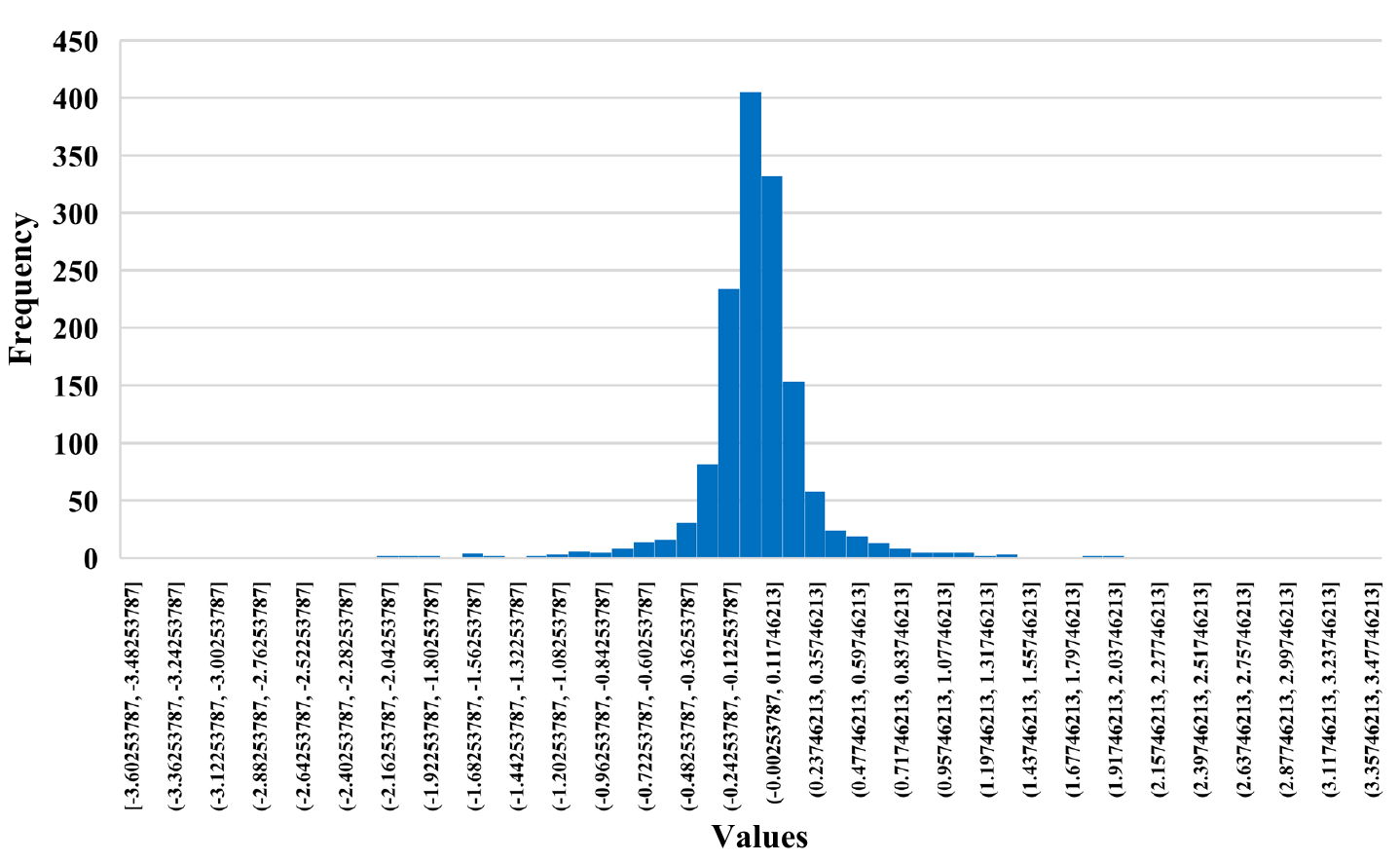}
\vspace{-0.4cm}
\caption{Probability distribution of latent representation \(y\).} 
\label{distribution}
\vspace{-0.4cm}
\end{figure}
\vspace{-0.08cm}
\subsubsection{Addition Branch}
AdderNet was first proposed by Chen \textit{et al.} \cite{AdderNet}, which utilizes \(L1\) norm rather than cross-correlation to calculate the similarities between inputs and filters. Due to the properties of the \(L1\) norm, AdderNet can avoid expensive multiplication and only involves addition or subtraction. What is more, the vanilla AdderNet adopts a batch normalization (BN) after each adder layer to prevent gradients from exploding. However, for the image compression task, it is not helpful to improve performance once introducing the BN layer since it cannot reduce pixel-wise redundancies. Thus, it is necessary to develop a new scheme that can not only prevent gradients from exploding, but also help to reduce spatial redundancies. 
\begin{figure*}[t]
\centering  
\includegraphics[width=\textwidth, height =7.2cm]{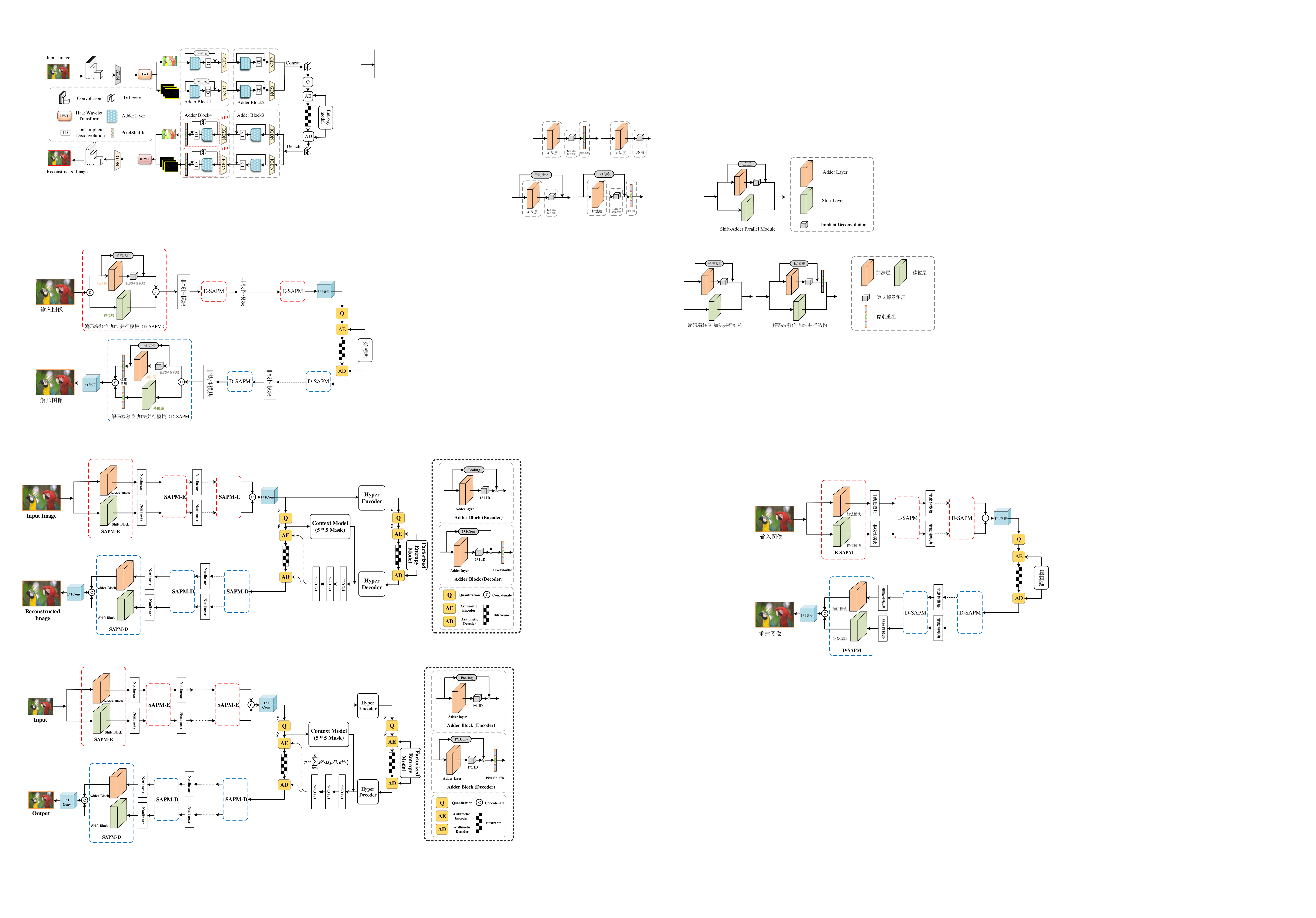}  
\vspace{-0.8cm}
\caption{General architecture. The detailed structures of the adder blocks are shown on the right, while the shift block is a shift layer in the encoder and a shift layer follow by a PixelShuffle layer \cite{Pixelshuffle} in the decoder. Context model is composed of a 5\(\times\)5 mask convolution \cite{mask}, while the structures of hyper encoder, hyper decoder and factorized entropy model are the same as \cite{Cheng2020}.} 
\label{framework}  
\vspace{-0.3cm}
\end{figure*}  
Motivated by \cite{deconvolution}, we adopt 1\(\times\)1 implicitly deconvolution (ID) to replace the BN after each adder layer. The core of it is to calculate the covariance matrix \(Cov\) of input \(X\) and then get the deconvolution operation \(D\):
\vspace{-0.1cm}
\begin{equation}
\operatorname{\textit{Cov}}=\frac{1}{N}(X-\mu)^{T}(X-\mu), D=\operatorname{\textit{Cov}}^{-\frac{1}{2}}
\vspace{-0.1cm}
\end{equation}
where \(N\) and \(\mu\) represent the number of samples and the mean of input, respectively. Once \(D\) is obtained, we can apply it after each centered input \((X-\mu).\) Furthermore, according to the association rule of matrix multiplication, we can implicitly obtain the output \(Y\) by changing the weights \(W\) of networks as follow:
\vspace{-0.1cm}
\begin{equation}
Y=(X-\mu) \cdot D \cdot W=X \cdot(D \cdot W)-\mu \cdot D \cdot W
\vspace{-0.1cm}
\end{equation}Although the computation cost of 1\(\times\)1 ID is slightly higher than the BN, it is still significantly lower than that of regular convolution layers (\textit{e.g.}, 5\(\times\)5 Conv).

Besides, two kinds of shortcut connections are developed for powerful expressive capacity, and the details will be discussed in Sec. \ref{overall structure}. Similar to CNN, the addition branch can extract fine-grained features and capture details in images.
% \vspace{-0.15cm}
\subsection{Laplace Mixture Model}
Entropy models utilize parameterized distribution to fit the marginal distribution of quantized latent representation \(\hat{y}\). In other words, a more accurate entropy model can save more bits, which can help to improve rate-distortion performance significantly. Ball\'e \textit{et al.} \cite{Balle2018} firstly proposed a univariate Gaussian distribution model for the hyperprior, and later works \cite{Minnen2018a} extended it to a mean and scale Gaussian distribution for accurate entropy estimation. Furthermore, Cheng \textit{et al.} \cite{Cheng2020} developed a Gaussian Mixture Model, which utilizes \(k\) Gaussian distributions to fit the marginal distribution and demonstrates strong fitting abilities. In a word, these works are based on an assumption that the prior distribution of latent representation is a Gaussian distribution. However, whether the prior distribution is also Gaussian in the proposed frameworks is unclear, and it should be explored.

We conduct several experiments and randomly select one of the channels to visualize the probability distribution of latent representation in Fig. \ref{distribution}. From the figure, we can see that the probability distribution of latent representation is more Laplace than Gaussian, which means that the entropy models used in CNN frameworks cannot fit the marginal distribution well. To this end, we propose Laplace Mixture Model (LMM) to estimate the marginal distribution more accurately.
\vspace{-0.2cm}
\begin{equation}
p_{\hat{y} \mid z}(\hat{y} \mid \hat{z}) \sim \sum_{k=1}^{K} w^{(k)} \mathcal{L}\left(\mu^{(k)}, \sigma^{(k)}\right)
\vspace{-0.2cm}
\end{equation}
where \(\mu^{(k)}\), \(\sigma^{(k)}\), \(w^{(k)}\)denote the mean, scale and weight of the \(k\)-th mixture (Laplace distribution) respectively. Once LMM is developed, it can help to save more bits during arithmetic encoding and decoding as shown in Fig. \ref{framework}

\subsection{Overall Structure}
\label{overall structure}
Fig. \ref{framework} shows the general architecture. The backbone network is based on \cite{Balle2017,Balle2018} and there are \(n\) levels of transformation in the encoder or decoder, where the convolution layer and transposed convolution layer are replaced by SAPM-E and SAPM-D, respectively. In SAPM-E, an adder block is composed of an adder layer followed by a 1\(\times\)1 ID. At the same time, we add an average pooling to improve the expressive capacity of the addition branch. In SAPM-D, 1\(\times\)1 ID is used to increase the number of output channels, and the upsampling can be realized by the PixelShuffle layer \cite{Pixelshuffle}. Similar to the role of pooling in the encoder, we also add a 1\(\times\)1 Conv as a shortcut connection to enhance the expressive capacity.

\section{Experiments}
\label{sec:Experiments}
\subsection{Experimental Setup}
\begin{figure*}[t]
\centering  
\includegraphics[width=\textwidth]{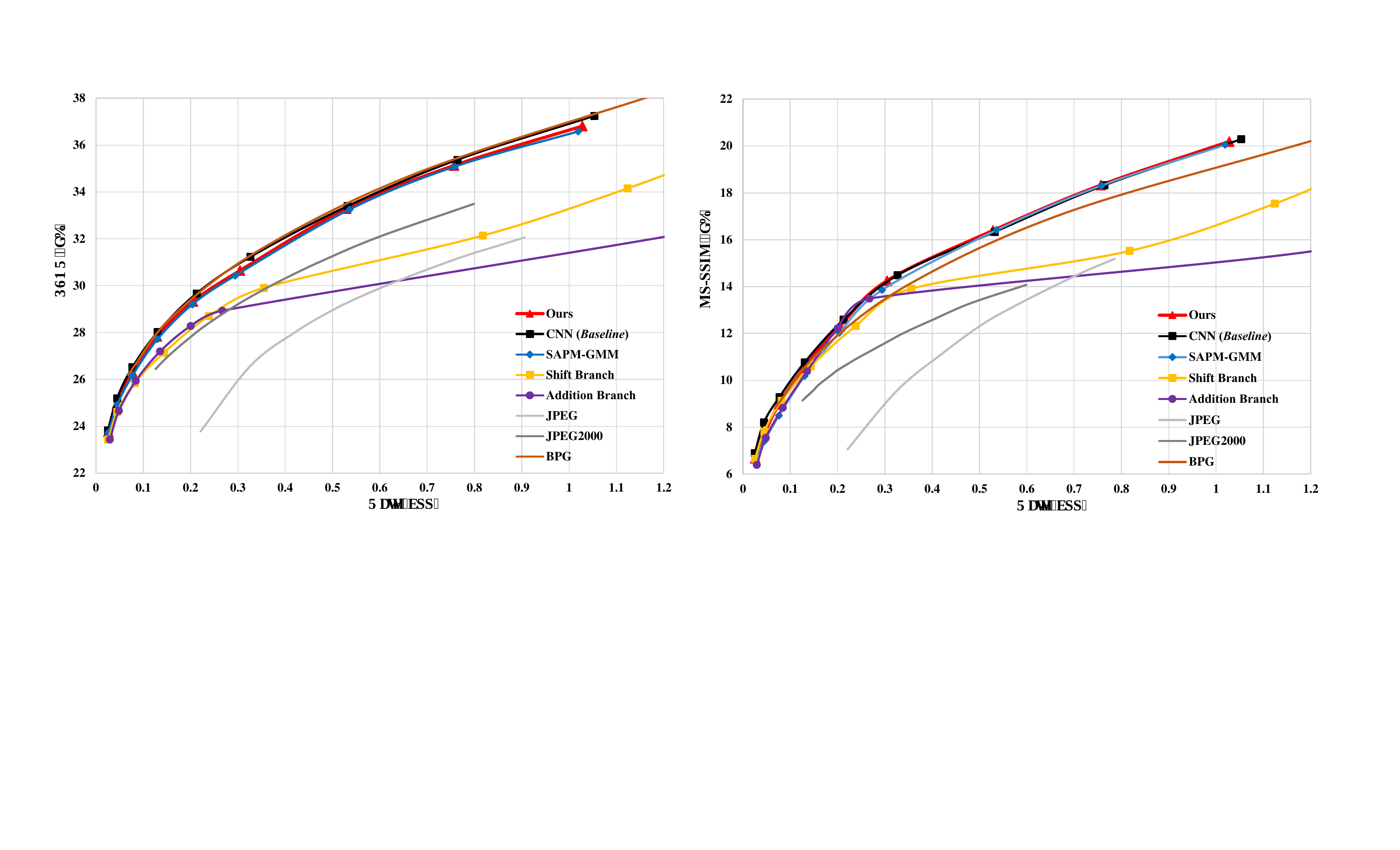}  
\vspace{-0.7cm}
\caption{Performance evaluation on Kodak dataset.} 
\label{RD}  
\vspace{-0.4cm}
\end{figure*}  
We adopt 3 levels of transformation in the encoder or decoder in our experiments. Specifically, for better performance, we employ a convolutional layer followed by two SAPM-Es in the encoder, and two SAPM-Ds followed by a transposed convolutional layer in the decoder, which is similar to the settings of vanilla AdderNet. Besides, we choose generalized divisive normalization (GDN) \cite{Balle2017} as our nonlinear module. Other settings such as hyper encoder, hyper decoder, and context model are consistent with \cite{Cheng2020}. After that, the model is trained and optimized for mean squared error (MSE) using two sets of \(\lambda\) values (\textit{i.e.,} low bitrate: \(\{16, 32, 64, 128, 256, 512\}\), high bitrate: \(\{1024, 2048, 4096\}\)) on the CLIC2020 training dataset \cite{CLIC2020} which consists of 61894 images with 256\(\times\)256 pixels, and tested on the standard Kodak dataset \cite{kodak} with 24 images of 512\(\times\)768 or 768\(\times\)512 pixels. To be specific, in the first stage, we train the models with the largest \(\lambda\) value in each set with a batch size of 8, and then apply Adam optimizer with the learning rate of \(1 \times 10^{-4}\) in the first 900,000 iterations and \(1 \times 10^{-5}\) in the remaining 100,000 iterations. For other bitrates (\(\lambda\) values), we adopt the models trained on bitrate \(\lambda=4096\) and \(\lambda=512\) as pre-trained models and then fine-tune the rest models for 320,000 iterations with a learning rate of \(1 \times 10^{-5}\). At last, we adopt average bit-per-pixel (bpp), average peak signal-to-noise ratio (PSNR), average multi-scale structural similarity (MS-SSIM)  as well as energy consumption as the metrics to evaluate the performance.
\vspace{-0.15cm}
\subsection{Performance Comparison}
The rate-distortion performance on the Kodak dataset is shown in Fig. \ref{RD}. We compare our method with the corresponding CNN baseline and several traditional standards, including JPEG \cite{JPEG}, JPEG2000 \cite{JPEG2000} and BPG \cite{BPG}. Regarding PSNR, the proposed method largely outperforms JPEG and JPEG2000, while showing extremely comparable performance with the CNN baseline and BPG. When comes to MS-SSIM, our model greatly surpasses JPEG, JPEG2000 and BPG, and even exhibits better performance than the CNN baseline at some high bitrate points. Moreover, similar to the CNN counterpart, our method also produces visually pleasant reconstructed images as shown in Fig. \ref{visualization}.

At last, we calculate the energy consumption of the proposed SAPMs according to Tab. \ref{table1}. Due to the low computation cost of shortcut connections and 1\(\times\)1 ID, we omit them and then make a comparison with the convolutional layer.
From Tab. \ref{SAPM energy}, the proposed SAPMs can bring more than 2\(\times\) energy reduction compared to the convolutional layer, which facilitates deployment on devices with limited resources.
\begin{table}[H]
\vspace{-0.4cm}
\caption{Unit energy comparison between CNN and SAPM.}
\label{SAPM energy}
\centering%把表居中
\begin{tabular}{ll|cc}%四个c代表该表一共四列，内容全部居中
\toprule%第一道横线
Operation                & Format & Convolution & SAPM \\
\midrule%第二道横线 
\# Mult.                  & FP32   &  \(3.70\times1\)   &  0 \\
\multirow{2}{*}{\# Add.}  & FP32   &  \(0.90\times1\)   &   \(0.90\times2\)   \\
                         & FIX32  &   0  &    \(0.10\times1\)  \\
\# Shift                 & FIX32  &   0  &  \(0.13\times1\)    \\ 
\midrule%第三道横线 
\multicolumn{2}{c|}{Total Energy Costs (\textit{pJ})}  &  \textbf{4.60}   &   \textbf{2.03 \textcolor{red}{(2.2\(\times\)\(\downarrow\))} }  \\
\bottomrule%第四道横线
\vspace{-0.6cm}
\end{tabular}
\end{table}
\subsection{Ablation Studies}
To verify the effectiveness of the proposed methods, several ablation experiments are carried out as shown in Fig. \ref{RD}. To begin with, either shift branch or addition branch is less capable compared to their CNN baseline, while integrating these two weak players can bring strong expressive capacity. Furthermore, the proposed LMM achieves a more accurate entropy estimation for the marginal distribution, which can save more bits and further improve the rate-distortion performance. What is more, it is worth noting that the vanilla AdderNet cannot be directly used in image compression because it would cause great performance degradation and its rate-distortion (RD) curve cannot be displayed in Fig. \ref{RD}.

\section{Conclusion}
\label{sec:Conclusion}
In this paper, we proposed SAPMs to largely reduce energy consumption in learned image compression. Besides, we thoroughly analyze the probability distribution of the latent representation and then develop a novel Laplace Mixture Model for more accurate entropy estimation. At last, several experiments are carried out to demonstrate that the proposed model can achieve comparable or even better performance on both PSNR and MS-SSIM metrics to that of the CNN counterpart while reducing more than 2\(\times\) energy consumption.

\bibliographystyle{IEEEbib}
\bibliography{strings,refs}

\end{document}